\documentclass[prd,aps,twocolumn,amsmath,amssymb,nofootinbib,preprintnumbers]
{revtex4}

\usepackage{graphicx}
\usepackage{dcolumn}
\usepackage{bm}
\usepackage{amsmath}
\usepackage{amsfonts}
\usepackage{euscript,bbm}
\usepackage{ifthen}
\usepackage{psfrag}
\usepackage{slashed}
\usepackage{color}

\renewcommand{\slash}{\displaystyle{\not}}

\def\ls{\mathrel{\lower4pt\vbox{\lineskip=0pt\baselineskip=0pt
           \hbox{$<$}\hbox{$\sim$}}}}
\def\gs{\mathrel{\lower4pt\vbox{\lineskip=0pt\baselineskip=0pt
           \hbox{$>$}\hbox{$\sim$}}}}
\def\drawbox#1#2{\hrule height#2pt

\hbox{\vrule width#2pt height#1pt \kern#1pt
              \vrule width#2pt}
              \hrule height#2pt}

\def\Asym#1#2{\vcenter{\vbox{\drawbox{#1}{#2}
              \kern-#2pt       
              \drawbox{#1}{#2}}}}

\newcommand{\be}{\begin{equation}}
\newcommand{\ee}{\end{equation}}
\newcommand{\bea}{\begin{eqnarray}}
\newcommand{\eea}{\end{eqnarray}}

\begin{document}

\title{keV Photon Emission from Light Nonthermal Dark Matter}

\author{Rouzbeh Allahverdi$^{1}$}
\author{Bhaskar Dutta$^{2}$}
\author{Yu Gao$^{2}$}

\affiliation{$^{1}$~Department of Physics and Astronomy, University of New Mexico, Albuquerque, NM 87131, USA \\
$^{2}$~Mitchell Institute of Fundamental Physics and Astronomy, Department of Physics and Astronomy, Texas A$\&$M University, College Station, TX 77843-4242, USA}

\begin{abstract}
We propose a possible explanation for the recent claim of an excess at 3.5 keV in the X-ray spectrum within a minimal extension of the standard model that explains dark matter and baryon abundance of the universe. The dark matter mass in this model is ${\cal O}({\rm GeV})$ and its relic density has a non-thermal origin. The model includes two colored scalars of ${\cal O}({\rm TeV})$ mass ($X_{1,2}$), and two singlet fermions that are almost degenerate in mass with the proton ($N_{1,2}$). The heavier fermion $N_2$ undergoes radiative decay to the lighter one $N_1$ that is absolutely stable. Radiative decay with a life time $\sim 10^{23}$ seconds can account for the claimed 3.5 keV line, which requires couplings $\sim 10^{-3}-10^{-1}$ between $X_{1,2}, ~ N_{1,2}$ and the up-type quarks. The model also gives rise to potentially detectable monojet, dijet, and monotop signals at the LHC.
\end{abstract}
MIFPA-14-10 \\ March, 2014
\maketitle

\section{INtroduction}
Recently XMM-Newton observatory has found an excess at 3.5 keV X-ray in the  spectrum of 73 galaxy clusters~\cite{bib:KeVExcess, Boyarsky:2014jta}.
If this excess persists 
such a photon emission can be the result of the late decay and/or annihilation of multi-keV mass dark matter, or decay of a metastable particle to daughter(s) with a keV mass spilting.
Dark matter induced 3.5 keV photon  excess has been recently studied in the scenario of extend neutrino sector~\cite{bib:KeVExcess,bib:OtherSterile}, the axion~\cite{pheno:axion} or its supersymmertic partner axino~\cite{pheno:axino}, string moduli~\cite{pheno:moduli}, and annihilation or decay via low energy effective operators~\cite{pheno:effective}. In this paper, we investigate the possibility of the decay of a light nonthermal dark matter, as proposed in previous work~\cite{Bhaskar1}, which connects the DM relic density to baryongenesis.

\section{The model}

The model is a minimal extension of the SM that includes iso-singlet color-triplet scalars $X_\alpha$ and SM singlet fermions $N_a$. The Lagrangian includes only renormalizable interactions and is given by
\bea \label{lagran}
{\cal L} & = & {\cal L}_{\rm SM} + {\cal L}_{\rm new} \, \nonumber \\
{\cal L}_{\rm new} & = & (\lambda_{\alpha a i} X^*_\alpha N_a u^c_{i} + \lambda^\prime_{\alpha i j} {X}_\alpha {d^c_i} d^c_j + {1 \over 2} M_a N_a N_a + {\rm h.c.})\,  \nonumber \\
& + & m^2_\alpha |X_\alpha|^2 + ({\rm kinetic ~ terms}) \, .
\eea
Here $1 \leq i,j \leq 3$ denote flavor indices (color indices are omitted for simplicity), and $\alpha$ and $a$ denote the number of $X$ and $N$ fields respectively.

The model with two copies of $X$ ($\alpha = 1,2$) and one copy of $N$ ($a=1$) can give rise to baryogenesis and a viable DM candidate~\cite{Bhaskar1}. This is the minimum requirement for generation of baryon asymmetry from the interference of tree-level and one-loop diagrams in $X_{1,2}$ decay. Moreover, the $N$ field can be the DM if its mass lies within the range $m_p - m_e \leq M_1 \leq m_P + m_e$ ($m_p$ and $m_e$ being the proton mass and the electron mass respectively). The conditions $M_1 \leq m_p + m_e$ and $m_p - m_e \leq M_1$ ensure that the decay channels $N \rightarrow p + e^{-} + {\bar \nu}_e , ~ {\bar p} + e^{+} + \nu_e$ and $p \rightarrow N + e^{+} + \nu_e$ are kinematically forbidden. The former implies stability of $N$, while the latter prevents catastrophic proton decay. Therefore, together they yield a suitable DM candidate whose stability is tied to that of the proton.

As discussed in~\cite{Bhaskar1}, obtaining the correct DM abundance in this model requires a non-thermal mechanism. This is because $N$ interactions with the up-type quarks bring it into thermal equilibrium with the plasma at temperatures as low as $T \sim M_1 \approx {\cal O}({\rm GeV})$. However, for $m_{1,2} \gtrsim {\cal}({\rm TeV})$, to be compatible with the LHC bounds, thermal freeze-out results in overabundance of $N$ according to the Lee-Weinberg bound~\cite{LW}. Late decay of a scalar field that reheats the universe to sub-GeV temperatures provides a suitable non-thermal alternative for producing the correct DM relic abundance. The late decay produces $X_{1,2}$ whose subsequent decay creates DM particles and generates baryon asymmetry of the universe. 

Such a common non-thermal origin for DM and baryogenesis, when combined with fact that $N$ is a viable DM candidate only if $M_1 \approx {\cal O}({\rm GeV})$, can provide a natural explanation of the baryon-DM coincidence puzzle~\cite{bdm} (for a detailed discussion, see~\cite{Bhaskar1}).

\section{$3.5$ keV photon for late decay}

Now consider the case with two copies of $N$ ($a=1,2$) with $m_p - m_e \leq M_{1,2} \leq m_p + m_e$. In this case, then lighter singlet (which we call $N_1$) will be absolutely stable. The other field $N_2$, which is heavier, is unstable but can have a very long life time. If $N_2$ is stable on cosmological time scales, then both $N_1$ and $N_2$ contribute to the DM relic abundance. As mentioned above, $N_1$ and $N_2$ are produced from the decay of $X_{1,2}$. Hence, for $\lambda_1 \sim \lambda_2$, they have comparable abundances and make similar contributions to the DM relic density.

\begin{figure}
\includegraphics[scale=0.5]{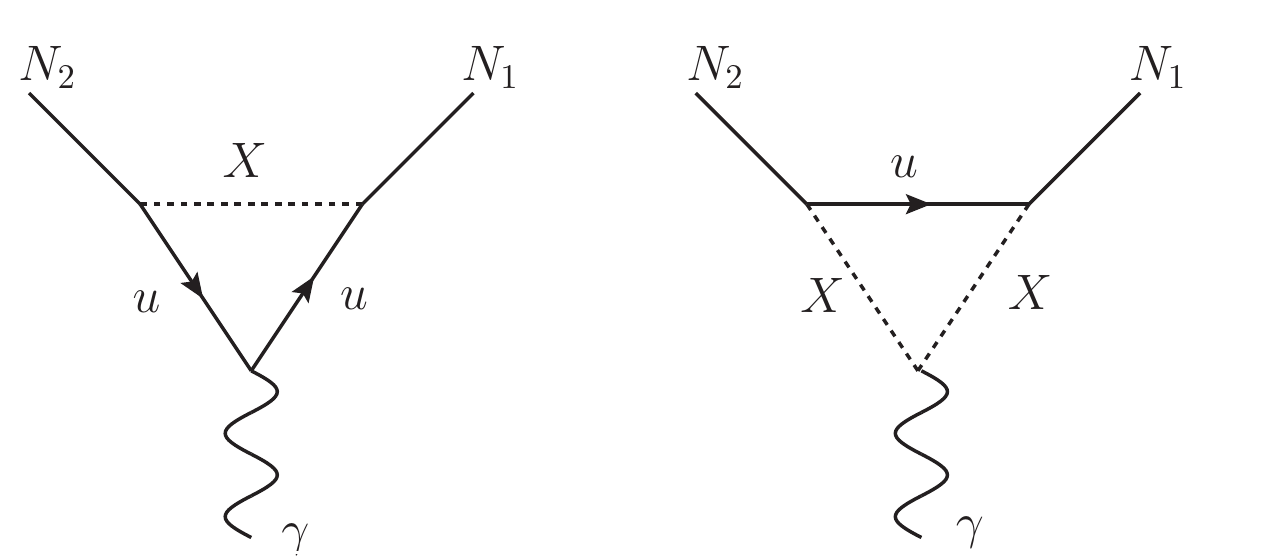}
\caption{Feynman diagrams for the $N_2\rightarrow N_1 \gamma$ decay.}
\label{fig:diagram}
\end{figure}

We note that $\Delta M \equiv M_2 - M_1 \leq 2 m_e$. Therefore, since both $N_1$ and $N_2$ are electrically neutral, the only decay mode that is allowed is $N_2 \rightarrow N_1 + \gamma$. The relevant diagrams for this decay are shown in Fig.~\ref{fig:diagram}. For an on-shell photon, gauge-invariance implies that only interactions terms of the form ${\bar \psi}_1 \sigma^{\mu \nu} F_{\mu \nu} \psi_{2}$ make a non-zero contribution to the decay process~\cite{LS}. Here $\psi_{1,2}$ are Majorana fermions that are made of Weyl fermions $N_{1,2}$.

The resulting decay width is given by
\begin{equation}
\Gamma_{N_2} \simeq 2^2\times{\vert \lambda_1 \lambda_2 \vert^2 \over 64 \pi^4} \alpha_{\rm em} {\Delta M}^3 {M^2_N \over m^4_X} .
\label{eq:widthLarge}
\end{equation}
Here we have assumed that $N_1,~N_2$ and $X_1,~X_2$ have similar masses that are denoted $M_N$ and $m_X$ respectively. Also, for simplicity, we have assumed that $X_1$ and $X_2$ have the same couplings, $\lambda_1$ and $\lambda_2$ respectively, to all flavors of $u^c$. With this assumptions, the factor of $2^2$ on the right-hand side of~(\ref{eq:widthLarge}) accounts for the interference of diagrams with $X_1$ and $X_2$ in the loop.

It is worth noticing that the decay width depends on the relative phase of $N_1$ and $N_2$ masses. As a result, a much narrower decay width can be achieved if $M_1$ and $M_2$ have opposite phases:
\begin{equation}
\Gamma_{N_2} \simeq 2^2 \times{\vert \lambda_1 \lambda_2 \vert^2 \over 64 \pi^4} \alpha_{\rm em} {{\Delta M}^5 \over m^4_X} .
\label{eq:widthSmall}
\end{equation}

To match the observed X-ray flux, the life time of $N_2$ should follow the estimate $\tau_{N_2} \approx 10^{29}~{\rm s} \cdot ({\rm keV}/M_N)$ according to Ref.~\cite{Boyarsky:2014jta}. The life time is inversely proportional to the DM mass for a given abundance in the host galaxy. Note that this estimate is subject to astrophysical uncertainties. For $M_N \approx {\cal O}({\rm GeV})$, we find $\Gamma_{N_2} =0.7 \times 10^{-47}$ GeV.
Considering $m_X \approx 2$ TeV (to be within the LHC reach) and $\Delta M = 3.5$ keV, we require $\vert \lambda_1 \lambda_2 \vert = 0.7 \times 10^{-6}$ for the generic case in Eq.~\ref{eq:widthLarge} and $\vert \lambda_1 \lambda_2 \vert = 0.2$ for the special case in Eq.~\ref{eq:widthSmall}, in order to account for the emission of 3.5 keV X-ray photon. Such values of $\lambda$ can give rise to successful baryogenesis (see~\cite{Bhaskar1}) as well as potential signals at the LHC (which we discuss in the next section).

\section{Other signals of the model}

Radiative decay $N_2 \rightarrow N_1 + \gamma$ is the most significant (and perhaps the only detectable) signal of the model in direct and indirect detection experiments. This decay can in principle produce a photon line with energy up to $2 m_e$, corresponding to the maximum value of $\Delta M$ allowed by stability of DM candidate and proton. In this case, however, we need much smaller coupling strengths than those mentioned in the previous section in order to have any observable signal at the present time.

Since $N$ only couples to a particular chirality of up-type quarks, spin-independent interactions that arise from effective operators of the form $(\bar \psi_N \psi_N) ({\bar \psi_q} \psi_q)/m^2_X$ between the DM and quarks are suppressed $\propto M^2_N/m^2_X$ ($\psi_N$ and $\psi_q$ are four-component fermions representing $N$ and quark fields respectively). There are other spin-independent interactions with nucleons that arise from twist-2 quark operators and one-loop diagrams that couple $N$ to gluons~\cite{DN}. The corresponding elastic scattering cross section for all of these interactions is suppressed $\propto m_X^{-8}$, which gives rise to $\sigma_{\rm SI} \ls 10^{-16}-10^{-15}$ pb for $m_X \sim {\cal O}({\rm TeV})$. This is considerably below the reach of upcoming experiments, and the neutrino background for DM mass of ${\cal O}({\rm GeV})$.

The spin-dependent cross-section is only suppressed $\propto m^{-4}_X$ as one has effective interactions of the form $({\bar \psi_N} \gamma_5 \gamma^\mu \psi_N) ({\bar \psi_q} \gamma_5 \gamma_\mu \psi_q)/m^2_X$. This results in $\sigma_{\rm SD} \ls 10^{-6}-10^{-5}$ pb, for $m_X \sim {\cal O}({\rm TeV})$, which is much below the bounds from current experiments~\cite{COUPP}, as well as the upcoming detection experiments. It is also significantly below the current LHC bounds on $\sigma_{\rm SD}$~\cite{Fox}, but in the vicinity of the LHC future reach~\cite{Tait}.

Indirect detection signals from DM annihilation will be negligible in this model. The DM annihilation rate is $\langle \sigma_{\rm ann} v \rangle \sim \lambda^4 \vert {\vec p} \vert^2/m^4_X$, where ${\vec p}$ is the three-momentum of annihilating DM particles. For $m_X \sim {\cal O}({\rm TeV})$ and $\lambda \leq {\cal O}(1)$, we have $\langle \sigma_{\rm ann} v \rangle \ll 10^{-31}$ cm$^3$ s$^{-1}$, which is well below the current limits from DM annihilation to photons set by the Fermi-LAT~\cite{Fermi}. The neutrino signal from annihilation of DM particles captured inside the Sun, which depends on $\sigma_{\rm SD}$, is also negligible due to its low capture rate, evaporation at $M_N\sim {\cal O}({\rm GeV})$ as well as incapability to annihilate into heavy mesons that can decay before being absorbed by solar medium.

DM interactions with matter have a novel signature in this model that may be seen. The effective interaction $N u^c d^c d^c$ leads to baryon destroying inelastic scattering of $N$ off nucleons, similar to the model in~\cite{Hylo}, which may have an appreciable rate for nucleon decay experiments.

Moreover, for $\lambda_1,\lambda_2 \lesssim 0.1$ this model can be probed at the LHC through its distinctive monojet signatures via a resonant $s$-channel exchange of $X$. If $\lambda_1,\lambda_2 \sim 10^{-3}$, the two jet + missing transverse energy ($\slash{E}_T$) channel can be searched for instead. The LHC search strategies have been studied in Ref.~\cite{Dutta:2014kia}. The coupling size required by the X-ray excess is below the current experimental constraint but can be confronted with future LHC data. If the $\lambda$ coupling to the top quark is significant, a monotop or ditop + $\slash{E}_T$ can also emerge.

\section{Conclusion}

In conclusion, we have shown that it is possible to explain the recent claim of an excess at 3.5 keV in the X-ray spectrum within a minimal extension of the SM that explains DM and baryon abundance of the universe from a non-thermal origin. The minimum field content that is required includes two colored scalars $X_{1,2}$ and two singlet fermions $N_{1,2}$. The $N_{1,2}$ fermions are almost degenerate in mass with the proton and are coupled to the up-type quarks through interaction terms $\lambda X^* N u^c$. The lighter singlet $N_1$ is absolutely stable, while the heavier one $N_2$ undergoes radiative decay $N_2 \rightarrow N_1 + \gamma$ with a long lifetime $\sim 10^{23}$ seconds. This decay produces the claimed 3.5 keV photon line for $m_X \sim {\cal O}({\rm TeV})$ and $\lambda \sim 10^{-3}$-$10^{-1}$. The model can also be probed through monojet, dijet, and monotop signals at the LHC.

\section{Acknowledgement}

The work of B.D. is supported in part by the DOE grant DOE Grant No. DE-FG02-13ER42020. Y.G. is supported by Mitchell Institute for Fundamental Physics and Astronomy. We also thank Louis Strigari for helpful communications.


\end{document}